\documentclass[fleqn, twocolumn, notitlepage, amsmath,amssymb,superscriptaddress]{revtex4}
\usepackage[colorlinks={true}, citecolor={blue}, filecolor={blue}, linkcolor={blue}, urlcolor={blue}]{hyperref}
\usepackage{epsfig}
\usepackage{amsmath}
\usepackage{graphicx}
\usepackage{color}
\usepackage{amsfonts,amssymb}
\usepackage[english]{babel}
\usepackage{hyperref}
\usepackage{lipsum}
\usepackage{graphics}
\usepackage{array}
\usepackage{tabu}
\usepackage{bm}
\usepackage{float}
\usepackage{setspace}

\newcommand{\be}{\begin{equation}}
\newcommand{\ee}{\end{equation}}
\newcommand{\ba}{\begin{eqnarray}}
\newcommand{\ea}{\end{eqnarray}}

\newcommand{\ket}[1]{|#1\rangle}

\newcommand{\abs}[1]{\left\vert#1\right\vert}

\begin{document}

\title{Ground-state cooling of mechanical resonators by quantum reservoir engineering}

\author{M. Tahir Naseem}
\affiliation{Department of Physics, Ko\c{c} University, 34450 Sariyer, Istanbul Turkey}

\author{\"{O}zg\"{u}r E. M\"{u}stecapl{\i}o\u{g}lu}
\email{omustecap@ku.edu.tr}
\affiliation{Department of Physics, Ko\c{c} University, 34450 Sariyer, Istanbul Turkey}

\begin{abstract}
{\bf Abstract.} 
Ground-state cooling of multiple mechanical resonators becomes vital to employ them in various applications ranging from ultra-precise sensing to quantum information processing. 
Here we propose a scheme for simultaneous cooling of multiple degenerate or near-degenerate mechanical resonators to their quantum ground-state, which is otherwise a challenging goal to achieve. As opposed to standard laser cooling schemes where coherence renders the motion of a resonator to its ground-state,  we consider an incoherent thermal source to achieve the same aim.  The underlying physical mechanism of cooling is explained by investigating a direct connection between the laser sideband cooling and ``cooling by heating". Our advantageous scheme of cooling enabled by quantum
reservoir engineering can be realized in various setups, employing parametric coupling of a cooling agent with the target systems. We also discuss using non-thermal baths to simulate ultra-high temperature thermal baths for cooling. 
 
\end{abstract}

\maketitle

\section*{Introduction}

The ground-state cooling of mechanical resonators is crucial for various applications, such as gravitational wave detection~\cite{Abramovici325}, ultrasensitive precision measurements~\cite{LaHaye74}, and for the investigation of the quantum behavior of mechanical systems~\cite{PhysRevLett.91.130401}, where only relevant motion is due to zero-point fluctuations. For the quantum information processing applications of optomechanical systems, mechanical resonators need to be close to their quantum ground-state~\cite{PhysRevLett.90.137901, PhysRevLett.98.030405}. To achieve this goal, different methods have been proposed and realized such as resolved sideband cooling~\cite{PhysRevLett.97.243905,PhysRevLett.99.093901,PhysRevLett.99.093902,PhysRevLett.99.137205,PhysRevA.77.033804,Schliesser2008,PhysRevA.79.039903,Teufel2011, Massel2012,PhysRevLett.116.063601,PhysRevLett.118.223602}, and feedback-assisted cooling~\cite{PhysRevLett.83.3174,Kleckner2006,PhysRevLett.99.160801,PhysRevLett.99.017201,Wilson2015,Rossi2018,PhysRevLett.123.203605,PhysRevResearch.2.033299}. In a standard approach, thermal energy is removed from a single mode of mechanical resonator to bring its state near to the ground-state. In the recent past, optomechanical systems having multiple mechanical resonators caught much attention~\cite{PhysRevLett.100.220401,PhysRevLett.110.253601,PhysRevLett.111.103605,PhysRevA.89.063805,PhysRevLett.112.014101}, finding various applications such as heat management~\cite{Xuereb_2015,PhysRevLett.120.060601} and mesoscopic entanglement~\cite{Ockeloen-Korppi2018}. However, ground-state cooling of degenerate or near-degenerate multiple mechanical resonators is quite a challenging task to accomplish in optomechanics. This is because the existence of dark modes in multiple resonators case, which decouple from the common cavity mode and hinders the simultaneous ground-state cooling of multiple resonators~\cite{Genes_2008, PhysRevLett.123.203605,PhysRevA.99.023826,PhysRevLett.112.013602}. There are few recent proposals to overcome this difficulty, among which two are prominent: (i) cooling based on dark-mode-breaking~\cite{PhysRevA.98.023860,PhysRevA.102.011502}, this demands phase-dependent coherent interaction between the mechanical resonators. (ii) Cold-damping feedback, in which input coherent cavity drive is dynamically modified using an electronic feedback loop~\cite{PhysRevResearch.2.033299,PhysRevLett.123.203605}. However, this technique does not work for the degenerate mechanical resonators and may not cool the resonators to their ground-state.

A complementary approach, namely ``cooling by heating'', exploits incoherent driving of the quantum system to cool it down~\cite{PhysRevLett.108.120602,PhysRevLett.108.120603}.  As opposed to laser cooling,  in this method, incoherent thermal drive removes energy from the quantum system only to dump it in another bath at a lower temperature. For optomechanical setting, this has been proposed for two optical modes coupled to a single mechanical resonator~\cite{PhysRevLett.108.120602}. 
This scheme requires a three-body interaction, and the 
the proposal is based on a mean-field approach in which correlations between the optical and mechanical modes are ignored, which is essentially valid in a very weak optomechanical coupling. Within the validity of this approach, it may not be possible to achieve ground-state cooling for experimentally realizable system parameters. In contrast, we propose a scheme beyond the mean-field approach and requires only two-body interaction to cool a mechanical resonator to its ground-state. We suggest electro-mechanical systems in which a superconducting microwave circuit is parametrically coupled to a mechanical resonator~\cite{Rocheleau2010,LaHaye2009,Connell2010,Teufel2011, Teufel2011b, Rodrigues2019} for the experimental realization of our proposal. We discuss the possibility of employing a squeezed thermal bath to cool mechanical resonators in setups that require an experimentally challenging ultra-high hot bath temperature for cooling.

In addition, simultaneous ground-state cooling of multiple degenerate or near-degenerate mechanical resonators is also possible, which is otherwise challenging to accomplish. A key to realizing this method is bath spectral filtering (BSF)~\cite{doi:10.1080/09500349414550381,PhysRevE.87.012140,PhysRevE.90.022102,Gelbwaser-Klimovsky2015,Ghosh12156,Ghosh9941,Naseem_2020, PhysRevResearch.2.033285}, in which unwanted frequencies are filtered out from the system-bath coupling spectra. The BSF has previously been shown to enhance different thermal functions~\cite{Naseem_2020, PhysRevResearch.2.033285}. We investigate a direct connection between the laser sideband cooling and cooling by heating in a standard optomechanical setting, which also explains the underlying physical cooling mechanism. In our present analysis, we propose a scheme for the simultaneous ground-state cooling of multiple degenerate or near-degenerate mechanical resonators. Replacing a laser drive with a spectrally filtered thermal bath or squeezed thermal bath enables simultaneous ground-state cooling, which is an otherwise challenging task.

\section*{Results}
%%%%%%%%%%%%%%%%%%%%%%%%%%%%%%%%%%%%%%%%%%% Figure 1 %%%%%%%%%%%%%%%%%%%%%%%%%%%%%%%%%%%%%%%%%%%%%%%%%%%
\begin{figure}[t]
\begin{center}
\includegraphics[width = 0.85\columnwidth]{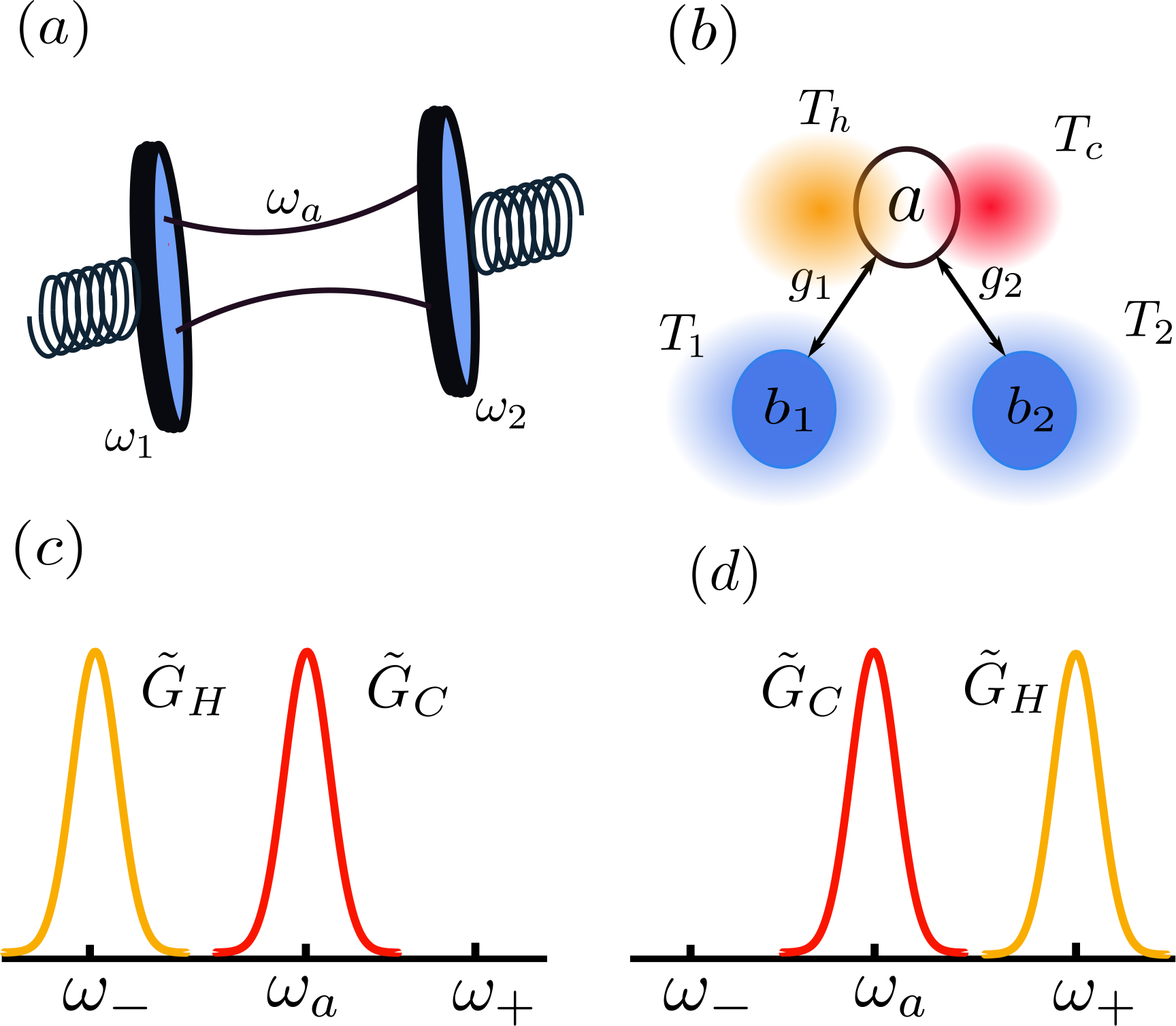}
\end{center}
\caption{{\bf System description.} (a) Schematic illustration of the optomechanical system being considered in this work. A single optical mode of frequency $\omega_{a}$ is coupled to two mechanical resonators of frequencies $\omega_{1}$ and $\omega_{2}$ via optomechanical couplings $g_{1}$ and $g_{2}$, respectively. (b) Each mechanical resonator ($b_{i}$) of frequency $\omega_{i}$ is in contact with an independent thermal bath at temperature $T_{i}$, $i=1, 2$. In addition, the optical mode is in contact with two independent thermal baths at temperatures $T_{h}$ and $T_{c}$. (c) and (d) filtered spectra of the optical hot and cold baths given by Eq.~(\ref{eq:FSB}), for cooling and heating process, respectively.}
\label{fig:1}
\end{figure}
%%%%%%%%%%%%%%%%%%%%%%%%%%%%%%%%%%%%%%%%%%%%%%%%%%%%%%%%%%%%%%%%%%%%%%%%%%%%%%%%%%%%%%%%%%%%%%%%%%%%%%%

{\bf Model.}~We consider a multimode optomechanical system (Fig.~\ref{fig:1}), which consists of a single optical mode interacting with two mechanical resonators. The optical mode is coupled to two thermal baths at temperatures $T_{c}$ and $T_{h}$, and each mechanical resonator of frequency $\omega_{i}$ is in contact with an independent thermal bath at temperature $T_{i}$, $i=1, 2$. The Hamiltonian of the isolated optomechanical system is given by (we take $\hbar =1$)
\begin{equation}\label{eq:OMS}
\hat{H}_{s} = \omega_{a}\hat{a}^{\dagger}\hat{a} + \sum^{2}_{i=1}\big[\omega_{i}\hat{b}^{\dagger}_{i}\hat{b}_{i} - g_{i}\hat{a}^{\dagger}\hat{a}(\hat{b}_{i}+\hat{b}^{\dagger}_{i})\big],
\end{equation}
where $\omega_{a}$ and $\omega_{i}$ are the frequencies of the optical and mechanical resonators, respectively. The bosonic annihilation (creation) operators of the optical and mechanical modes are given by $\hat{a}$ ($\hat{a}^{\dagger}$) and $\hat{b}_{i}$ ($\hat{b}^{\dagger}_{i}$), respectively. The $g_{i}$ terms represent the single-photon optomechanical coupling between the optical mode and mechanical resonators. This multimode optomechanical model can be realized via a photonic crystal optomechanical system~\cite{Fang2017} or circuit electromechanical system~\cite{Massel2011, Massel2012}.

The master equation in the interaction picture is given by
\begin{eqnarray}\label{eq:master}
&\dot{\tilde{\rho}}& =  \tilde{\mathcal{L}}_{H} + \tilde{ \mathcal{L}}_{C} + \tilde{\mathcal{L}}_{i},
\end{eqnarray}
where the interaction of the baths with the system is described by Liouville superoperators $\tilde{\mathcal{L}}_{x}$ ~\cite{PhysRevA.98.052123, Naseem_2020, PhysRevResearch.2.033285} (see Methods ``The master equation'').

For the numerical simulations, we have considered parameters of a typical optomechanical system~\cite{RevModPhys.86.1391}: $\omega_{i} = 2\pi\times 10$ MHz, $g_{i} = 2\pi\times 100$ kHz, $\kappa_{i} = 2\pi\times 100$ Hz, $\kappa_{h} = \kappa_{c} = 2\pi\times 1$ MHz.  Initial average phonons in the mechanical resonator is $\bar{n}_{i}(0)\sim 2\times10^3$, and the corresponding temperature becomes $T_{i}\sim 1.5$ K. We take the temperature of the cold bath $T_{c}\sim$ mK, while the temperature of the hot bath is larger by many order of magnitude, i.e., $4\times 10^4\lesssim T_{h}\lesssim 4\times 10^7$ K. In the results, we have scaled all these parameters with the optical frequency $\omega_{a}$.

{\bf Single mechanical resonator.}~
%%%%%%%%%%%%%%%%%%%%%%%%%%%%%%%%%%%% Figure 2 %%%%%%%%%%%%%%%%%%%%%%%%%%%%%%%%%%%%%%%%%%%%%%%%%%
\begin{figure}[t]
\centering
\includegraphics[width=0.78\columnwidth]{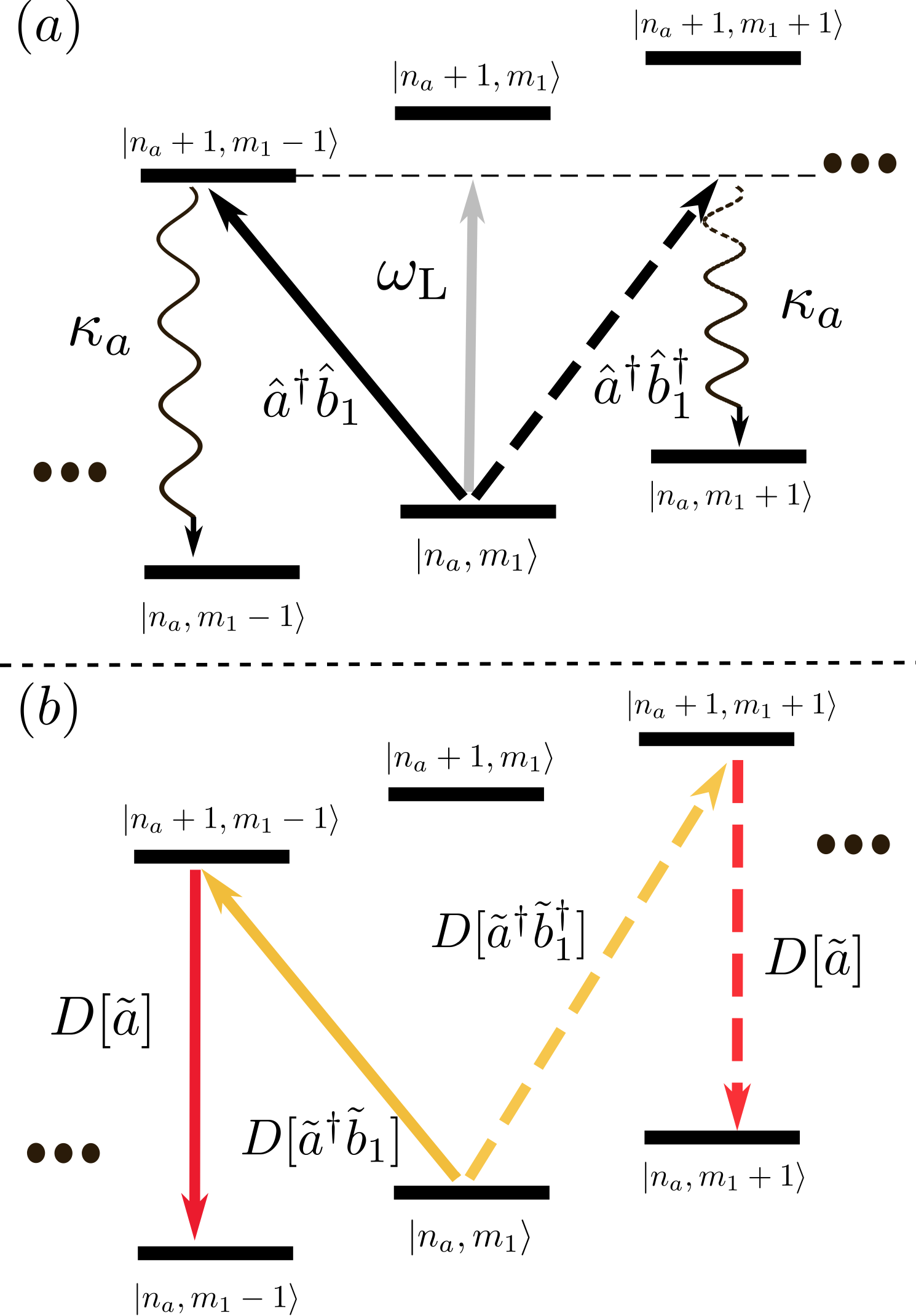}
\caption{\label{fig:2}{\bf Cooling mechanism.}  The underlying energy transitions between the dressed states $\ket{n_{a}, m_{1}}$ of the optomechanical system with a single mechanical resonator. In both panels, solid and dashed lines show the cooling and heating process, respectively. (a) The cooling and heating processes in the presence of coherent laser drive of frequency $\omega_\text{L}$. The cooling process proceeds by the absorption on the lower sideband, leading to the leakage of a blue-shifted photon from the cavity by destroying one phonon. On contrary, absorption at the upper sideband leads to the heating of the mechanical resonator~~\cite{Schliesser2008}.  
(b) The energy transitions involved in the cooling and heating process in our proposed scheme. As opposed to laser cooling, the transition at lower sideband ($\omega_{-}$) is induced by the hot bath via the incoherent system-bath interaction, given by $D[\tilde{a}\tilde{b}^{\dagger}_{1}]$ in Eq.~(\ref{eq:L_L}). This creates an energetic (blue-shifted) photon which is dumped in a bath at lower temperature $T_{c}$. Cooling with a single optical bath is not allowed by the second law of thermodynamics. Because, in the case of a single optical bath, $\ket{n_{a},m_{1}}\to\ket{n_{a}+1,m_{1}-1}\to\ket{n_{a},m_{1}-1}$ transitions mean heat flows from the mechanical to optical bath for $T_{1}<T_{h}$, hence we need another bath to be consistent with the second law of thermodynamics.
In the heating process, the hot bath induce the transition of frequency $\omega_{+}$ following by dumping of a photon in the cold bath, which results in heating of the mechanical resonator.
For the cooling process to dominate, the upper sideband frequency $\omega_{+}$ is filtered from the system-bath coupling spectra, as shown in Fig.~\ref{fig:1}(c).}
\end{figure}
%%%%%%%%%%%%%%%%%%%%%%%%%%%%%%%%%%%%%%%%%%%%%%%%%%%%%%%%%%%%%%%%%%%%%%%%%%%%%%%%%%%%%%%%%%%%%%%%
To explain the underlying physical mechanism, first, we will consider a single mechanical resonator. This simple model helps us to make a direct connection between the ``cooling by heating'' with laser sideband cooling obvious in the optomechanical system. In the case of a single mechanical resonator, the index $i$ can be dropped in Eq.~(\ref{eq:master}). We are interested in the dynamics of the mechanical resonator, therefore, upon taking trace over the optical mode, Eq.~(\ref{eq:master}) takes the form
\begin{eqnarray}\label{eq:mastertrace}
\frac{d\tilde{\rho}}{dt} = (A^{-}+A^{-}_{\text{th}})\tilde{\mathcal{D}}[\tilde{b}_{1}]+ (A^{+}+ A^{+}_{\text{th}})\mathcal{\tilde{D}}[\tilde{b}^{\dagger}_{1}],
\end{eqnarray}
here $A^{-}$ ($A^{-}_{\text{th}}$) is downward transition rate, and $A^{+}$ ($A^{+}_{\text{th}}$) is upward transition rate in the mechanical resonator due to the optical (mechanical) baths, and these are defined as
\begin{eqnarray}\label{eq:drift}
A^{-} &:=& \zeta^2_{1}\sum_{\alpha=H,C}(G_{\alpha}(\omega_{+})\langle\tilde{n}_{a}\rangle + G_{\alpha}(-\omega_{-})\langle\tilde{n}_{a}+1\rangle),\nonumber \\
A^{+} &:=& \zeta^2_{1}\sum_{\alpha=H,C}\big(G_{\alpha}(\omega_{-})\langle\tilde{n}_{a}\rangle + G_\alpha(-\omega_{+})\langle\tilde{n}_{a}+1\rangle\big),\nonumber \\
A^{-}_{\text{th}} &:=& G_{1}(\omega_{1}), \qquad A^{+}_{\text{th}} := G_{1}(-\omega_{1}).
\end{eqnarray}
Note that we use $A^{\pm}$ for the transition rates to compare our results with the sideband cooling in the atomic and optomechanical systems~\cite{RevModPhys.86.1391}.

In the rest of the work, we consider one dimensional Ohmic spectral densities of all the baths, given by
~\cite{PhysRevE.90.022102, Gelbwaser-Klimovsky2015, PhysRevE.87.012140}
\begin{eqnarray}\label{eq:SRF}
G_{x}(\omega)=
\begin{cases}
\kappa_{x}(\omega)[1 + \bar{n}_{x}(\omega)] &\omega> 0, \\
\kappa_{x}(\abs{\omega})\bar{n}_{x}(\abs{\omega}) &\omega< 0, 
\end{cases}
\end{eqnarray}
here, $\kappa_{x}$ is the system-bath coupling strength, and $\bar{n}_x (\omega)= 1/[e^{(\omega/T_{x})} - 1]$ is the mean number of quanta in the respective bath.
For the cooling of mechanical resonator, bath spectral filtering is applied only to optical baths, which yields~\cite{doi:10.1080/09500349414550381,PhysRevResearch.2.033285} 
\begin{eqnarray}\label{eq:FSB}
\tilde{G}_\alpha := \frac{1}{\pi} \frac{\pi G_\alpha(\omega)}{[(\omega - (\omega(n+1/2)+\Delta_\alpha(\omega))^2
+(\pi G_\alpha(\omega))^2]},\nonumber\\
\end{eqnarray}  
here $\Delta_\alpha(\omega)$ is the Lamb-shift induced by the thermal bath.

In a typical optomechanical system, resolved-sideband ($\omega_{1}>\kappa_{h,c}$) cooling can be induced if the input laser light of frequency $\omega_{L}$ is resonant with the lower sideband $\omega_\text{L} = \omega_{-} = \omega_{a}-\omega_{1}$~\cite{Teufel2011, RevModPhys.86.1391}. In this case, the Stokes process is suppressed and anti-Stokes scattering dominates, as shown in Fig.~\ref{fig:2}(a). As long as the upper sideband $\omega_{+} = \omega_{a}+\omega_{1}$ is far off-resonant, the input laser can drive the mechanical resonator near to its ground-state. Compared with the standard optomechanical setup, we don't have input laser drive contribution in the Hamiltonian given in Eq.~(\ref{eq:OMS}). Instead, the cavity is driven by the hot and cold thermal baths, however, both lower and upper phonon sidebands appear as a result of these thermal drives. By analogy with laser sideband cooling, we may want to drive the cavity at the lower sideband and suppress the upper sideband. This can be achieved by using bath spectral filtering~\cite{doi:10.1080/09500349414550381,PhysRevE.87.012140,PhysRevE.90.022102,Gelbwaser-Klimovsky2015,Ghosh12156,Ghosh9941,Naseem_2020, PhysRevResearch.2.033285}.

To drive the cavity at lower sideband with the hot thermal bath, the filtered hot bath couples only to a transition frequency at $\omega_{-}$ (lower sideband), and the coupling frequency at upper sideband $\omega_{+}$ is filtered (Fig.~\ref{fig:1}(c)). To complete the refrigeration process, the cold bath couples to a transition frequency at $\omega_{a}$. The coupling spectra of the hot and cold baths are filtered and well-separated (Fig.~\ref{fig:1}(c)), and satisfy
\begin{eqnarray}\label{eq:Rspectrum}
\tilde{G}_C (\omega_a)&\gg&  \tilde{G}_H (\omega_a),\\ \nonumber
\tilde{G}_C (\omega_{-})&\ll& \tilde{G}_H (\omega_{-}).
\end{eqnarray}
For the choice of this bath spectral filtering,  
the master equation given in Eq.~(\ref{eq:master}) modifies to
\begin{eqnarray}
\tilde{\mathcal{L}}_{C} &=& \tilde{G}_{C}(\omega_{a})\tilde{\mathcal{D}}[\tilde{a}]
	+ \tilde{G}_{C}(-\omega_{a})\tilde{\mathcal{D}}[\tilde{a}^{\dagger}],\nonumber \\ 	
\tilde{\mathcal{L}}_{H}	&=& \zeta_{1}^2\big(\tilde{G}_{H}(\omega_{-})\tilde{\mathcal{D}}[\tilde{a}\tilde{b}^{\dagger}_{1}]
	+ \tilde{G}_{H}(-\omega_{-})\tilde{\mathcal{D}}[\tilde{a}^{\dagger}\tilde{b}_{1}]\big),\nonumber \\ 
	\tilde{\mathcal{L}}_{1}&=& G_{1}(\omega_{1})\tilde{\mathcal{D}}[\tilde{b}_{1}]
	+ G_{1}(-\omega_{1})\tilde{\mathcal{D}}[\tilde{b}_{1}^{\dagger}],\label{eq:filtMEM}
\end{eqnarray}
we note that $\tilde{\mathcal{L}}_{1}$ remains the same. 
%%%%%%%%%%%%%%%%%%%%%%%%%%%%%%%%%%%% Figure 3 %%%%%%%%%%%%%%%%%%%%%%%%%%%%%%%%%%%%%%%%%%%%%%%%%%
\begin{figure}[t]
\centering
\includegraphics[width=0.50\textwidth]{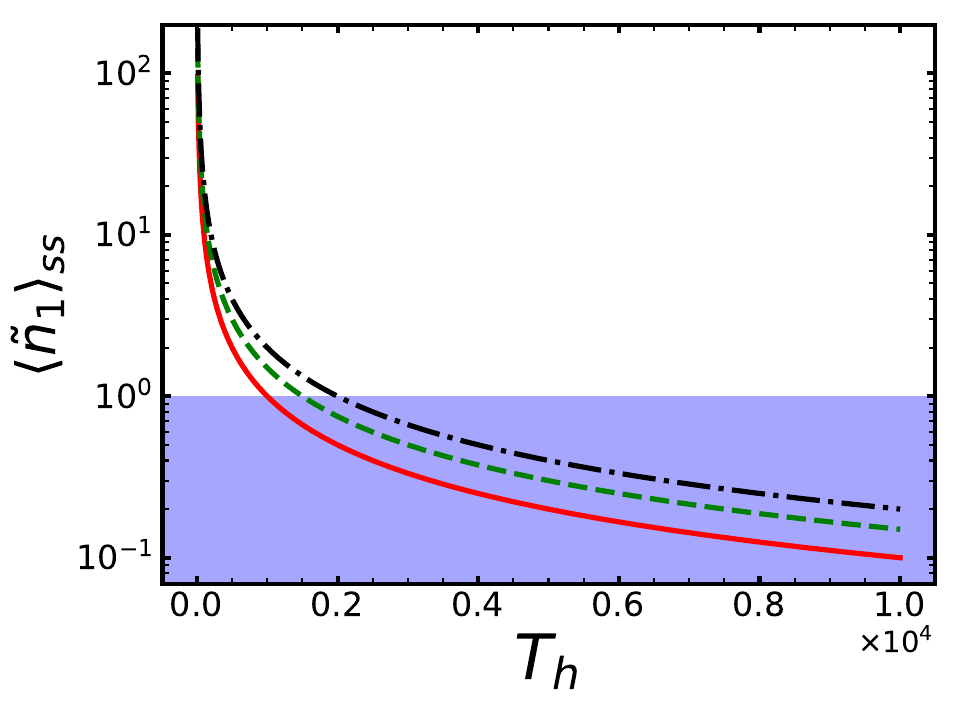}
\caption{\label{fig:3}{\bf Single mechanical resonator cooling.}  The steady-state mean phonon number $\langle \tilde{n}_{1}\rangle_{ss}$ as a function of the hot bath temperature $T_{h}$. Solid, dashed, and dot-dashed lines are for $T_{1} = 1\times 10^{-4}, 2\times10^{-4}$ and $3\times10^{-4}$, respectively. The other parameters are $\omega_{a} = 1$, $\omega_{1} = 1\times 10^{-7}$, $\kappa_{h} = \kappa_{c} = 1\times 10^{-8}$, $\kappa_{1} = 1\times 10^{-12}$, $T_{c}=1\times 10^{-5}$, and $g_{1} = 1\times 10^{-9}$. All the parameters are scaled with the optical frequency $\omega_{a} = 2\pi \times 10^{14}$ Hz. The temperatures in SI units are given as $T_{c} \approx 5$ mK, $T_{1} \approx 0.5, 1$ and $1.5$ K for solid, dashed, dot-dashed lines, respectively. The hot bath temperature is higher by many order of magnitude and it is approximately $4\times 10^4\lesssim T_{h}\lesssim 4\times 10^7$ K.}
\end{figure}
%%%%%%%%%%%%%%%%%%%%%%%%%%%%%%%%%%%%%%%%%%%%%%%%%%%%%%%%%%%%%%%%%%%%%%%%%%%%%%%%%%%%%%%%%%%%%%%%

%%%%%%%%%%%%%%%%%%%%%%%%%%%%%%%%%%%%%% Figure 4 %%%%%%%%%%%%%%%%%%%%%%%%%%%%%%%%%%%%%%%%%%%%%%%%%%%%%%%
\begin{figure}[t]
\centering
\hspace*{-0.2cm}\includegraphics[width=0.50\textwidth]{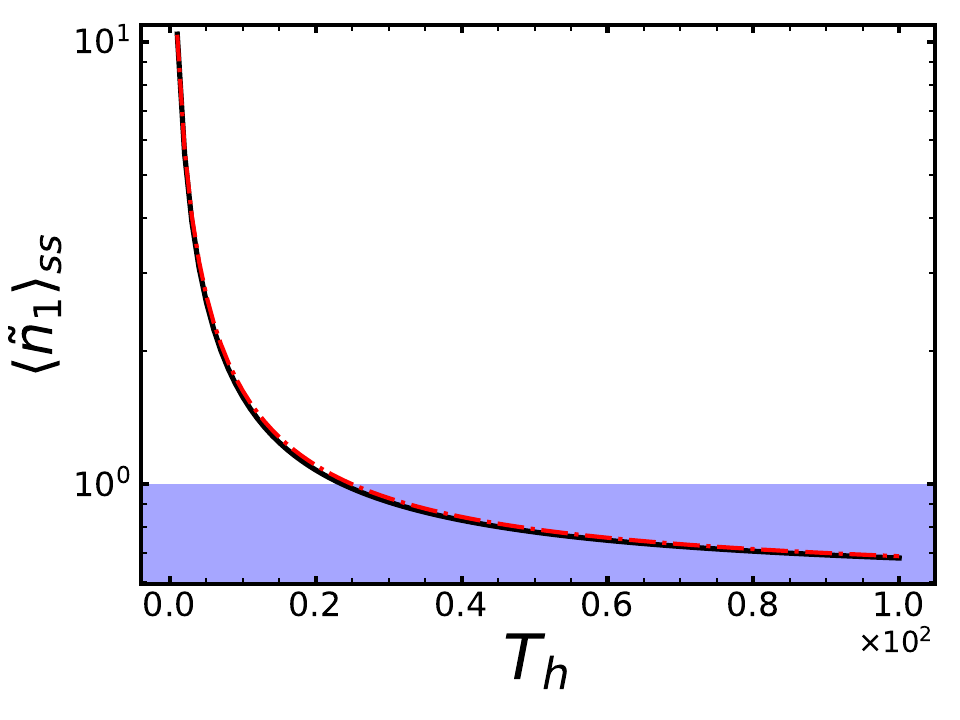}
\caption{\label{fig:4}{\bf Numerical vs approximate mean phonon number.} The steady-state mean phonon number $\langle \tilde{n}_{1}\rangle_{ss}$ as a function of the hot bath temperature $T_{h}$. Solid line is for
Eq.~(\ref{eq:nbs}) in case of ignoring $\zeta_{1}^2$ terms from $\langle \tilde{n}_{a}\rangle_{ss}$ (Eq.~(\ref{eq:appna})), and dot-dashed line represents $\langle \tilde{n}_{1}\rangle_{ss}$ calculated by the numerical simulation of the master equation given in Eq.~(\ref{eq:filtMEM}). In the numerical simulation, we use truncated Hilbert space with 7 photon states and 70 phonon states. The use of small Hilbert space dimensions for the optical mode is justified, because $\langle\tilde{n}_{a}\rangle_\text{ss}\ll 1$ for all parameters considered. We have verified that the increase in the Hilbert space dimensions does not affect the results.
Parameters: $\bar{n}_{1}=10$, $\bar{n}_{c}=0.5$, $\omega_{a} = 1$, $\omega_{1} = 1\times 10^{-7}$, $\kappa_{h} = \kappa_{c} = 1\times 10^{-8}$, $\kappa_{1} = 1\times 10^{-12}$, and $g_{1} = 1\times 10^{-9}$. All the parameters are scaled with the optical frequency $\omega_{a} = 2\pi \times 10^{14}$ Hz. %and the corresponding bath temperatures are $T_{1}\approx 47$ mK, $T_{c}\approx 4800$ K,
}
\end{figure}
%%%%%%%%%%%%%%%%%%%%%%%%%%%%%%%%%%%%%%%%%%%%%%%%%%%%%%%%%%%%%%%%%%%%%%%%%%%%%%%%%%%%%%%%%%%%%%%%%%%%%

The optical mode reaches to almost a thermal steady-state at temperature $T_{c}$ with corrections of the order $\mathcal{O}(\zeta^{2})$~\cite{Gelbwaser-Klimovsky2015,Ghosh12156,Latune2019}, and contribution from the mechanical bath. In our results, we ignore the correction term of the order $\zeta^{2}$ from the steady-state mean photon number $\langle\tilde{n}_{a}\rangle_\text{ss}$, which then takes the form
\begin{equation}\label{eq:appna}
\langle\tilde{n}_{a}\rangle_\text{ss} \approx \frac{\tilde{G}_{C}(-\omega_{a})+\tilde{G}_{1}(-\omega_{1})}{\big(\tilde{G}_{C}(\omega_{a})+\tilde{G}_{C}(-\omega_{a})\big)+\big(\tilde{G}_{1}(\omega_{1})+\tilde{G}_{1}(-\omega_{1})\big)}.
\end{equation}
The comparison between the numerical simulation of the master equation given in Eq.~(\ref{eq:filtMEM}), and our approximate results is presented in Fig~\ref{fig:4}.
From Eq.~(\ref{eq:filtMEM}), one can obtain the mean number of phonons in the mechanical resonator
\begin{eqnarray}
\frac{d\langle\tilde{n}_{1}\rangle}{dt} = (A_{c}^{+}+A^{+}_\text{th})\langle\tilde{n}_{1}+1\rangle - (A_{c}^{-}+A^{-}_\text{th})\langle\tilde{n}_{1}\rangle, 
\end{eqnarray}
here $A_{c}^{+}$ ($A_{c}^{-}$) represents upward (downward) transition rate due to the optical filtered baths (Fig.~\ref{fig:1}(c)), and given by
\begin{eqnarray}
A_{c}^{-}  &:=& \zeta_{1}^2\tilde{G}_{H}(-\omega_{-})\langle\tilde{n}_{a}+1\rangle, \quad
A_{c}^{+} := {\zeta}_{1}^2\tilde{G}_{H}(\omega_{-})\langle\tilde{n}_{a}\rangle .\nonumber
\end{eqnarray} 
The steady-state average phonon number is
\begin{eqnarray}\label{eq:nbs}
\langle\tilde{n}_{1}\rangle_{\text{ss}} = \frac{A^{+}_{c}+\kappa_{1}\bar{n}_{1}}{\tilde{\Gamma}_{c} + \kappa_{1}},
\end{eqnarray}
here $\tilde{\Gamma}_{c}=A^{-}_{c}-A^{+}_{c}$ is the net optical baths damping rate.
This equation of phonon-number is similar to the steady-state phonon number in a standard optomechanical system for laser sideband cooling~\cite{PhysRevLett.99.093902}, which is given by $\langle n_{1}\rangle_{ss} = (A^{+}+\kappa_{1}\bar{n}_{1})/(\Gamma_\text{opt}+\kappa_{1})$. %Here, 
%$\bar{n}^{o}_{1}$ is the quantum back-action limit~\cite{PhysRevLett.99.093902}, 
$\Gamma_\text{opt}=A^{-}-A^{+}$ is the net optomechanical damping rate, further, $A^{+}$ and $A^{-}$ represent the rates of Stokes (heating) and anti-Stokes (cooling) processes, respectively. 
For the resolved-sideband %($\omega_{1}>\kappa_{h,c}$)%
ground-state laser cooling, (i) upper sideband should be far off-resonant, (ii) the optical damping $\Gamma_\text{opt}$ must be much greater than the mechanical damping $\kappa_{1}$, and (iii) the initial phonon number $\bar{n}_{1}(0)$ must be much smaller than the quality factor of the mechanical resonator $\bar{n}_{1}\ll\omega_{1}/\kappa_{1}$~\cite{PhysRevLett.99.093902}. 
Similar to the resolved-sideband cooling, in our scheme ground-state cooling is possible provided, (i) upper sideband ($\omega_{+}$) from the hot bath spectrum is filtered,
(ii) for sufficiently low cold bath temperature, the thermal dissipation due to the optical baths is greater than the mechanical bath coupling strength, i.e.,
$\tilde{\Gamma}_{c}\gg\kappa_{1}$. (iii) The initial phonon number in the mechanical resonator is much smaller than its quality factor $\bar{n}_{1}\ll\omega_{1}/\kappa_{1}$, in addition, we note that these conditions are stated for $T_{h}>T_{1}>T_{c}$.
For the bath spectra shown in Fig.\ref{fig:1}(c), and considered system parameters (Fig.~\ref{fig:3}), $\tilde{\Gamma}_{c}$ is larger than $\kappa_{1}$ by three order of magnitude. The other conditions mentioned here also satisfy with in the considered system parameters. To investigate the steady-state cooling performance of our scheme, we plot the average number of phonons $\langle\tilde{n}_{1}\rangle_{ss}$ in the mechanical resonator as a function of the hot bath temperature $T_{h}$ in Fig.~\ref{fig:3}. The results indicate that ground-state cooling is possible for ultra-high hot bath temperatures. We remark that the mirror has no access to the ultra-high temperature bath. It only couples to the optical cavity field, which is in a single-mode thermal state. The only interaction with the single-mode cavity field and the mirror is through the optomechanical coupling. The molecular or atomic vibrational modes of the mirror have no resonant interaction with the cavity mode. At all times, the mirror only evolves towards its center-of-mass motional ground state.

The cooling mechanism can be explained by direct analogy with the laser sideband cooling shown in Fig.~\ref{fig:2}. By increasing the temperature $T_{h}$ of the hot bath, the number of photons of ``correct frequency" in the cavity to induce the red (lower) sideband transition increases. In a refrigeration process, the hot bath induces a transition of frequency $\omega_{-}$ at the cost of destroying a single phonon (Fig.~\ref{fig:2}). Then the energetic (blue-shifted) photon created in previous transition
is dumped in to the cold bath to complete the refrigeration process. Note that, this explanation is only valid for the filtered baths spectra shown in Fig.~\ref{fig:1}(c). For baths spectra including upper phonon sideband or spectrally overlapping baths, this explanation is invalid.

If we select the optical baths spectra of the form shown in Fig.~\ref{fig:1}(d), and they follow
\begin{eqnarray}\label{eq:Rspectrum}
\tilde{G}_C (\omega_a)&\gg&  \tilde{G}_H (\omega_a),\\ \nonumber
\tilde{G}_C (\omega_{+})&\ll& \tilde{G}_H (\omega_{+}),
\end{eqnarray}
then the master equation given in Eq.~(\ref{eq:master}) simplifies to
\begin{eqnarray}
\label{eq:filtMEH2}
\tilde{\mathcal{L}}_{C} &=& \tilde{G}_{C}(\omega_{a})\tilde{\mathcal{D}}[\tilde{a}]
	+ \tilde{G}_{C}(-\omega_{a})\tilde{\mathcal{D}}[\tilde{a}^{\dagger}],\\ 	
\tilde{\mathcal{L}}_{H}	&=& \zeta_{1}^2\big(\tilde{G}_{H}(\omega_{+})\tilde{\mathcal{D}}[\tilde{a}\tilde{b}_{1}]
	+ \tilde{G}_{H}(-\omega_{+})\tilde{\mathcal{D}}[\tilde{a}^{\dagger}\tilde{b}^{\dagger}_{1}]\big),\label{eq:filtMEC2}
\end{eqnarray}
and $\tilde{\mathcal{L}}_{1}$ remains the same. The energy transitions induced by these dissipators ($\tilde{\mathcal{D}}[\tilde{a}^{\dagger}\tilde{b}^{\dagger}], \tilde{\mathcal{D}}[\tilde{a}\tilde{b}]$) lead to heating of the mechanical resonator as shown in Fig.~\ref{fig:2}. $\Gamma_{h}=A^{-}_{h}-A^{+}_{h}$ is the net optical thermal dissipation rate with the downward and upward transition rates
\begin{eqnarray}
A^{-}_{h}  &:=& \zeta_{1}^2\tilde{G}_{H}(\omega_{+})\langle\tilde{n}_{a}\rangle,\quad
A^{+}_{h} := {\zeta}_{1}^2\tilde{G}_{H}(-\omega_{+})\langle\tilde{n}_{a}+1\rangle. \nonumber
\end{eqnarray}
The dynamical evolution of the initial phonon number is given by
\begin{equation}
\langle\tilde{n}_{1}(t)\rangle= \frac{A^{+}_{h}+\kappa_{1}\bar{n}_{1}}{\tilde{\Gamma}_{h} +\kappa_{1}} + \bar{n}_{1}e^{-t(\tilde{\Gamma}_{h}+\kappa_{1})},
\end{equation} 
in this case, $\tilde{\Gamma}_{h}$ is negative, consequently, the mechanical resonator heats up and the system becomes unstable in the long-time limit

{\bf Multiple mechanical resonators.} Here, we consider two mechanical resonators couple to a single optical cavity, as shown in Fig.~\ref{fig:1}, however, the extension of our scheme to N mechanical resonators is straightforward.
For the case of the hot and cold baths spectra considered in Fig.~\ref{fig:1}(c), the reduced master equation of the two mechanical resonators simplifies to
\begin{eqnarray}\label{eq:mastertrace2}
\frac{d\tilde{\rho}}{dt} = (A_{i}^{-}+A^{-}_{\text{th},i})\tilde{\mathcal{D}}[\tilde{b}_{i}]+ (A_{i}^{+}+ A^{+}_{\text{th},i})\mathcal{\tilde{D}}[\tilde{b}^{\dagger}_{i}],
\end{eqnarray}
and the transitions rates due to the optical baths have the form
\begin{eqnarray}
A_{i}^{-}  &:=& \zeta_{i}^2\tilde{G}_{H}(-\omega_{-,i})\langle\tilde{n}_{a}+1\rangle, \quad
A_{i}^{+} := {\zeta}_{i}^2\tilde{G}_{H}(\omega_{-,i})\langle\tilde{n}_{a}\rangle , \nonumber
\end{eqnarray} 
$\omega_{-, i} = \omega_{a} - \omega_{i}$.

%%%%%%%%%%%%%%%%%%%%%%%%%%%%%%%%%%%% Figure 5 %%%%%%%%%%%%%%%%%%%%%%%%%%%%%%%%%%%%%%%%%%%%%%%%%%
\begin{figure}[t]
\centering
\includegraphics[width=0.50\textwidth]{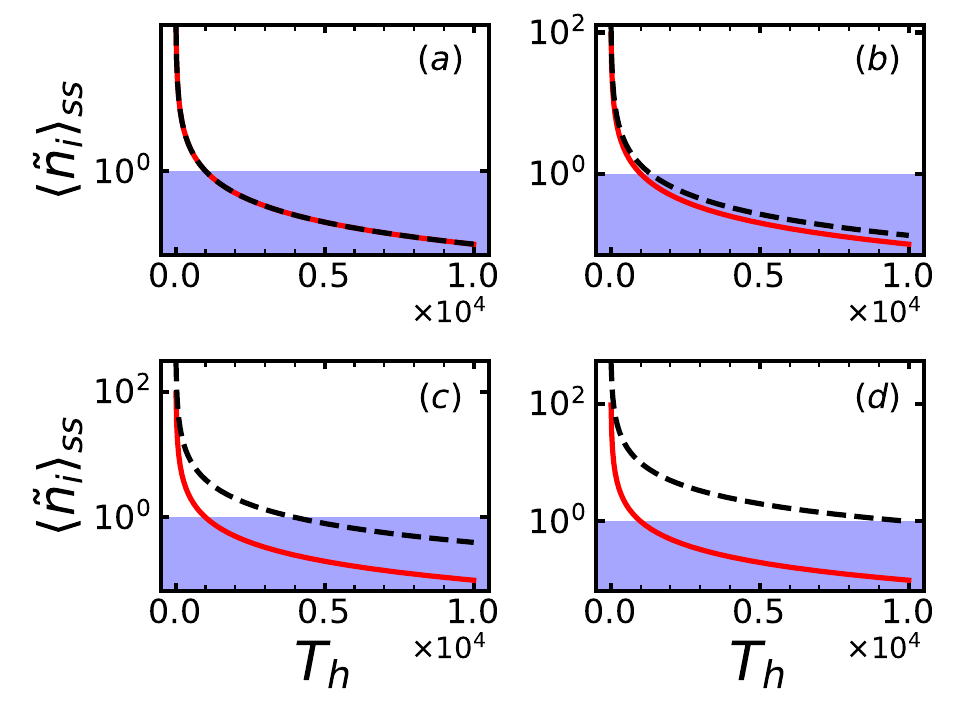}
\caption{\label{fig:5}{\bf Multiple mechanical resonators cooling.} The steady-state mean phonon number $\langle \tilde{n}_{i}\rangle_{ss}$ as a function of the hot bath temperature $T_{h}$. Solid, and dashed lines are for the first and second mechanical resonator, respectively. Parameters: $T_{1}=T_{2} = 2\times10^{-4}$, $T_{c}=1\times 10^{-5}$, $\omega_{a} = 1$, $\omega_{1} = 1\times 10^{-7}$, $\kappa_{h} = \kappa_{c} = 1\times 10^{-8}$, $\kappa_{1} = 1\times 10^{-12}$, and $g_{1} = 1\times 10^{-9}$. All the parameters are scaled with the optical frequency $\omega_{a} = 2\pi \times 10^{14}$ Hz. (a) $\omega_{1}=\omega_{2}$, $g_{1} = g_{2}$, (b) $\omega_{2}=0.75$ $\omega_{1}$, $g_{1}= g_{2}$,  (c) $g_{2} = 0.5$ $g_{1}$, $\omega_{1}=\omega_{2}$, and (d) $\kappa_{2} = 10$ $\kappa_{1}$, $\omega_{1}=\omega_{2}$, $g_{1}=g_{2}$.}
\end{figure}
%%%%%%%%%%%%%%%%%%%%%%%%%%%%%%%%%%%%%%%%%%%%%%%%%%%%%%%%%%%%%%%%%%%%%%%%%%%%%%%%%%%%%%%%%%%%%%%%

At steady-state the average phonon number in each resonator is given by
\begin{eqnarray}\label{eq:nbs2}
\langle\tilde{n}_{i}\rangle_{\text{ss}} = \frac{A^{+}_{i}+\kappa_{i}\bar{n}_{i}}{\tilde{\Gamma}_{i} + \kappa_{i}},
\end{eqnarray}
here $\tilde{\Gamma}_{i}=A_{i}^{-}-A_{i}^{+}$. This shows that the steady-state mean phonon number of each resonator depends on the relaxation rates of both resonators with their baths. The simultaneous cooling of both degenerate and non-degenerate multiple resonators is possible provided these relaxation rates are not very high. 
On contrary, in optomechanical sideband cooling, existence of the dark modes hinder the simultaneous multiple resonators cooling~\cite{PhysRevA.99.023826}. In addition, for degenerate or nearly degenerate resonators, one can not cool down either of the mechanical resonators~\cite{Genes_2008, PhysRevLett.123.203605}. In our scheme, Eq.~(\ref{eq:nbs2}) shows that, simultaneous cooling of both degenerate and non-degenerate resonators is possible. This is shown in Fig.~\ref{fig:5}, in which we plot the steady-state phonon number $\langle\tilde{n}_{i}\rangle_{ss}$ of each resonator as a function of the hot bath temperature $T_{h}$. For the identical resonators (Fig.~\ref{fig:5}(a)) coupled to thermal baths of the same temperature, the cooling curves are identical. However, for the non-degenerate case (Fig.~\ref{fig:5}(b)), the steady-state phonons in the second resonator is greater than the first resonator. The lesser cooling of the second resonator is because of the choice of the hot bath spectrum (Fig.~\ref{fig:1}(c)), which peaks at $\omega_{a}-\omega_{1}$. The second resonator frequency $\omega_{2}$ is below this peak, due to which it is weakly driven as compared to the resonator one, consequently, $\tilde{\Gamma}_{2}<\tilde{\Gamma}_{1}$. The effect of optomechanical coupling strength $g_{i}$ is shown in Fig.~\ref{fig:5}(c), the weakly coupled second resonator have smaller optical thermal dissipation ($\tilde{\Gamma}_{2}<\tilde{\Gamma}_{1}$), which results in lesser cooling. The mechanical resonator with larger system-bath coupling cools lesser as shown in Fig.~\ref{fig:5}(d).

Extension of our scheme to N independent mechanical resonators coupled to a single cavity mode is straightforward. Because Eqs.~(\ref{eq:mastertrace2}) and (\ref{eq:nbs2}) remain valid for arbitrary number of independent resonators. Another key feature of our method is that it is not only limited to optomechanics, but can also be realized in different platforms provided the target system to be cooled interacts dispersively with the working medium through the Hamiltonian of the form $\hat{H}_\text{int}=gN_{o}X_{m}$. Where $X_{m}$ is observable of the target system, and $N_{o}=\eta\hat{H}_\text{wm}$ with $\eta$ a positive constant, such an interaction yields a master equation of the form given in Eq.~(\ref{eq:master}). For example, a qubit dispersively coupled to a superconducting resonator~\cite{PhysRevE.90.022102, RevModPhys.85.623}, or with a
nanomechanical resonator~\cite{Armour_2008,LaHaye2009,Connell2010}.

{\bf Cooling by heating via a non-thermal bath.}
In the sideband cooling, the anti-Stokes (heating) process competes with the Stokes (cooling) process. In cavity optomechanics, if a cavity is coherently driven at the lower sideband, this leads to the suppression of Stokes scattering rate (Fig.~\ref{fig:2}(a)), which cools the motion of a mechanical resonator. However, due to the vacuum fluctuations, the Stokes heating processes are not completely suppressed, which sets a %fundamental
 limit to achieve true ground-state cooling~\cite{PhysRevLett.99.093902}. For the optimal set of the system parameters, one can achieve the quantum back-action limit~\cite{PhysRevLett.99.093902}, which is not a fundamental limit. There are several proposals to cool a mechanical resonator below this limit~\cite{PhysRevLett.102.207209,PhysRevLett.108.153601,PhysRevA.90.013824}. It has been proposed that squeezed light can theoretically eliminate the quantum back-action limit~\cite{PhysRevA.94.051801}, and the experiment showed a significant advantage of using squeezed light over laser cooling with coherent light~\cite{Clark2017}. Similarly, squeezed thermal reservoirs show %an
advantages over thermal baths by simultaneously increasing the cooling power and efficiency of quantum absorption refrigerators~\cite{Correa2014}.

 %%%%%%%%%%%%%%%%%%%%%%%%%%%%%%%%%%%% Figure 6 %%%%%%%%%%%%%%%%%%%%%%%%%%%%%%%%%%%%%%%%%%%%%%%%%%
\begin{figure}[t]
\centering
\includegraphics[width=0.50\textwidth]{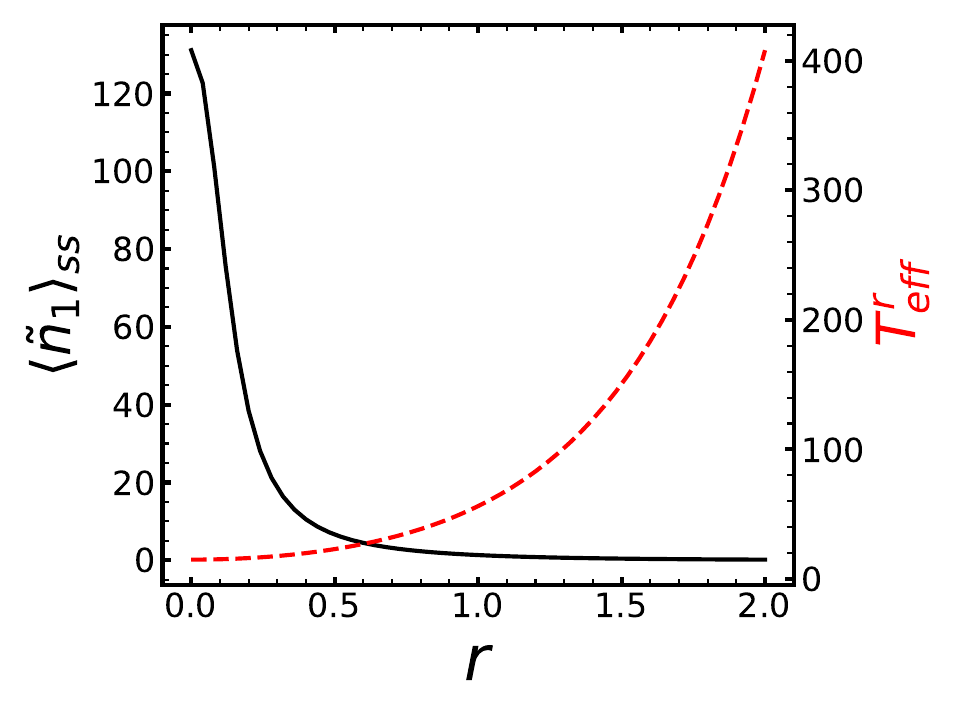}
\caption{\label{fig:6}{\bf Cooling assisted by a squeezed thermal bath.} The steady-state mean phonon number $\langle \tilde{n}_{1}\rangle_{ss}$, and effective temperature $T^{r}_\text{eff}$ of the squeezed thermal bath as a function of the squeezing parameter r. Solid, and dashed lines are for mean phonon number $\langle \tilde{n}_{1}\rangle_{ss}$ and effective temperature $T_\text{eff}$, respectively. Parameters are $\omega_{a} = 2\pi\times 7$ GHz, $\omega_{1} = 2\pi\times 10$ MHz, $g_{1} = 2\pi\times 100$ kHz, $\kappa_{c} = 2\pi\times 1$ MHz, $\tilde{\kappa}_{h} = 2\pi\times 1.5$ MHz, $\kappa_{1} = 2\pi\times 32$ Hz, and $T_{1}=T_{c}\approx 65$ mK. For $\langle \tilde{n}_{1}\rangle_{ss}$, all the parameters are scaled with the microwave resonator frequency $\omega_{a} = 2\pi \times 7$ GHz, and the squeezed thermal bath effective temperature $T^{r}_\text{eff}$ is calculated in SI units for $\bar{n}_{h}=0$.}
\end{figure}
%%%%%%%%%%%%%%%%%%%%%%%%%%%%%%%%%%%%%%%%%%%%%%%%%%%%%%%%%%%%%%%%%%%%%%%%%%%%%%%%%%%%%%%%%%%%%%%%
%%%%%%%%%%%%%%%%%%%%%%%%%%%%%%%%%%%% Figure 7 %%%%%%%%%%%%%%%%%%%%%%%%%%%%%%%%%%%%%%%%%%%%%%%%%%
\begin{figure*}[!t]
\centering
\includegraphics[width=0.85\textwidth]{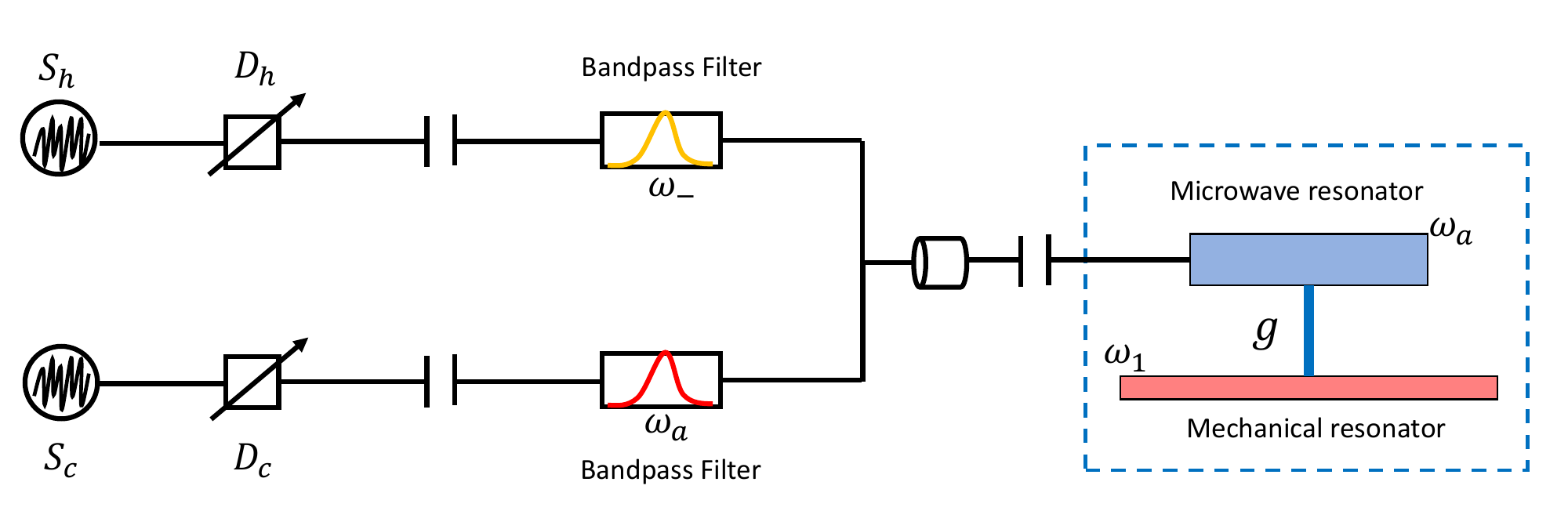}
\caption{\label{fig:7}{\bf Experimental feasibility.} Realization of the proposed cooling scheme employing an electro-mechanical system, an analog of the optomechanical system in Fig.~\ref{fig:1}. In this system, a mechanical resonator of frequency $\omega_{1}$ is parametrically coupled (with a coupling strength $g$) to a superconducting microwave resonator~\cite{Rocheleau2010} of frequency $\omega_{a}$ or a qubit~\cite{Connell2010, LaHaye2009} or a microwave circuit~\cite{Teufel2011b,Teufel2011, Rodrigues2019} and the system is placed inside a dilution refrigerator~\cite{Rocheleau2010,Teufel2011}. The multimode circuit electro-mechanical system can be considered for simultaneous cooling of the multiple mechanical resonators~\cite{Massel2012}. The microwave resonator is driven by carefully filtered hot and cold white noise sources $S_c$, and $S_h$, respectively. The amplitude of these drives is controlled by attenuators $D_c$ and $D_h$, respectively. The white noise sources can be realized by considering copper thin-film resistors~\cite{Senior2020}, or arbitrary waveform generators can be used to produce noisy electric fields to drive the electrical component of the electro-mechanical system, and a hot bath of very high-temperature $T_h\sim9500$ K can be engineered~\cite{PhysRevX.7.031044}. The bandpass filters which yield effective Lorentzian form spectra of the cold and hot baths centered at frequency $\omega_{a}$, and $\omega_{-}$ can be implemented by coupling two microwave resonators designed at frequency $\omega_{a}$, and $\omega_{-}$, respectively~\cite{PhysRevA.90.032114, Senior2020}. %for the cold and hot baths can be implemented by coupling two microwave resonators designed at frequency $\omega_{a}$, and $\omega_{-}$, respectively~\cite{Senior2020}. It yields effective Lorentzian form spectra of the cold and hot baths centered at frequency $\omega_{a}$, and $\omega_{-}$, respectively~\cite{PhysRevA.90.032114}.
 In a non-thermal bath induced cooling, the white noise source $S_h$ can be replaced by a squeezed electronic noise source~\cite{PhysRevX.7.031044}. For a very large squeezing parameter, the squeezed electronic noise source can be realized by considering an on-chip micromaser setup, where the interactions between a two-level system and a microwave resonator are tunable and can be turned on and off repeatedly~\cite{PhysRevA.81.052121}. %The microwave resonator of micromaser designed at frequency $\omega_{-}$ that is directly coupled to the electro-mechanical system's electrical component can serve for spectral filtering~\cite{Senior2020}.
}
\end{figure*}
%%%%%%%%%%%%%%%%%%%%%%%%%%%%%%%%%%%%%%%%%%%%%%%%%%%%%%%%%%%%%%%%%%%%%%%%%%%%%%%%%%%%%%%%%%%%%%%%

We discussed the possibility of ground-state cooling by using a spectrally filtered ultra-hot thermal bath (see Results ``Single mechanical resonator'' ). It can be challenging to realize a bath of such a high temperature. To overcome this challenge, we propose to use a squeezed thermal bath to cool mechanical resonators. %We replace the hot thermal bath with a squeezed thermal bath in the scheme proposed in Sec.~\ref{sec:model}.
 We consider an electro-mechanical system~\cite{Rocheleau2010,Teufel2011, Teufel2011b} with a single mechanical resonator due to the numerical computational constraints of considering a large Hilbert space size. By the replacement of hot thermal bath with a squeezed thermal bath in the scheme proposed in Fig.~\ref{fig:1}, the dissipator $\tilde{\mathcal{L}}_{h}$ given in Eq.~(\ref{eq:filtMEM}) takes the form~\cite{breuer2002}
\begin{eqnarray}\label{eq:squeezdissp}
\tilde{\mathcal{L}}^{r}_{H}	&=& \tilde{\kappa}_{h}\zeta_{1}^2\big((N+1)\tilde{\mathcal{D}}[\tilde{w}] + N\tilde{\mathcal{D}}[\tilde{w}^{\dagger}]\big)\nonumber \\
	&-& \tilde{\kappa}_{h}\zeta_{1}^2M\tilde{\mathcal{D}}[\tilde{w},\tilde{w}] -\tilde{\kappa}_{h}\zeta_{1}^2M\tilde{\mathcal{D}}[\tilde{w}^{\dagger},\tilde{w}^{\dagger}],
\end{eqnarray}
where $\tilde{w}=\tilde{a}\tilde{b}^{\dagger}$, $\tilde{\mathcal{D}}[A,B]:=2A\tilde{\rho}B-BA\tilde{\rho}-AB\tilde{\rho}$, and $\tilde{\kappa}_{h}$ is the effective system-bath coupling after applying bath spectral filtering~\cite{Senior2020}.  For the squeezed thermal bath, we have assumed that the jump operators related with the decay processes into channels of frequencies $\pm\omega_{-}$ only are non-zero.
Rest of the dissipators associated with the cold and mechanical bath in Eq.~(\ref{eq:filtMEM}) remain the same. The squeezing dependent coefficients $N$ and $M$ are given by
\begin{eqnarray}
N &:=& \bar{n}_{h}(\text{cosh}^2r + \text{sinh}^2r) + \text{sinh}^2r \\
M &:=& -\text{sinh}\,r\:\text{cosh}\,r\,(2\bar{n}_{h}+1),
\end{eqnarray}
where $r$  is the squeezing parameter, and we have set the squeezing phase to zero. A unitary transformation can always be found to cast the dissipators in Eq.~(\ref{eq:squeezdissp}) into a standard Lindblad form of Eq.~(\ref{eq:filtMEM}). This ensures that the system's effective dynamics remain completely positive and trace-preserving despite a squeezed thermal drive~\cite{Lindblad1976}. The steady-state mean phonon number shown in Fig.~\ref{fig:6} is obtained by a numerical solution of Eq.~(\ref{eq:filtMEM}) by replacing $\tilde{\mathcal{L}}_{H}$ with Eq.~(\ref{eq:squeezdissp}). In the numerical simulation, we used truncated Hilbert space with four-photon states and 400 phonon states. %The use of small Hilbert space dimensions for electromagnetic mode is justified because $\langle n_{a}\rangle_\text{ss}\ll 1$ for the set of parameters considered in Fig.~\ref{fig:6}.
 The squeezing dependent effective temperature of the squeezed thermal bath can be evaluated by~\cite{PhysRevE.86.051105}
\begin{equation}
T^{r}_\text{eff} = \frac{\hbar\omega_{-}}{k_B\text{log}[\frac{\text{tanh}^2r+e^{\hbar\omega_{-}/k_{B}T_{h}}}{1+\text{tanh}^2r e^{\hbar\omega_{-}/k_{B}T_{h}}}]}\geq T_{h}.
\end{equation}
The results in Fig.~\ref{fig:6} show that ground-state cooling is possible for a set of experimentally realizable system parameters, including squeezing parameter $r\sim1.2$~\cite{PhysRevLett.104.251102}, and much higher squeezing parameters are possible by using a non-thermal coherent atomic reservoir~\cite{Entropy2016}.

For implementing our scheme in an optical regime with a squeezed thermal reservoir, merging the optomechanical system described in Fig.~\ref{fig:1} with the micromaser scheme of heating~\cite{PhysRev.159.208} can be an appealing possibility. In this setup, the cold bath could be an electromagnetic vacuum that is always and unavoidably coupled to the cavity mode. Such a cold bath, %spectral filtering is not required, because
 typically has a Lorentzian spectrum~\cite{RevModPhys.86.1391}, which can be controlled by cavity linewidth~\cite{Gelbwaser-Klimovsky2015}. The hot bath can be realized by sending a beam of atoms prepared in quantum coherent superpositions that interact with the cavity mode by the Tavis-Cummings model~\cite{Entropy2016,Turk2020}. In this setup, the hot bath spectral filtering is not required if the atoms are resonant with the lower sideband of the cavity.
In addition, a very large squeezing parameter $r\sim 3.5$ can be achieved~\cite{Entropy2016}. The micromaser setup can be extended beyond the beam of atoms to on-chip systems, where the interactions between the two-level system and the resonator are tunable and can be turned on and off repeatedly~\cite{PhysRevA.81.052121}. More efficient non-thermal quantum thermalizing effective baths can be designed by injecting quantum coherence between atomic levels~\cite{Entropy2016,Turk2020}. The on-chip micromaser setup can be coupled with an electro-mechanical system to implement our cooling scheme (see Fig.~\ref{fig:7} and Discussion ``Experimental Feasibility'' for details). Similar to such a micromaser setting, a periodical drive and Floquet method to get the same effect of a very hot bath is also possible~\cite{PhysRevE.85.061126}.

\section*{Discussion} 

Our advantageous scheme for ground-state cooling of multiple mechanical resonators enabled by bath spectral filtering can be experimentally realized in various setups that may employ analogs of an optomechanical system described in Fig.~\ref{fig:1}. 

{\bf Experimental Feasibility.} (A) {\color{blue}In the previous section}, we have argued that it is possible to cool the motion of single or multiple mechanical resonators to their quantum ground-state using spectrally filtered incoherent light, as shown in Fig.~\ref{fig:4}. However, this requires an ultra-high temperature of the hot bath, which can be challenging to realize in the current state-of-the-art experimental setups. To overcome this difficulty, here we suggest an experimental setup based on an electro-mechanical system~\cite{Rocheleau2010,Teufel2011, Teufel2011b}, which can be used to implement our proposal. The setup is shown in Fig.~\ref{fig:7}, in which a micromechanical resonator of frequency $\omega_{1}$ is parametrically coupled to a superconducting microwave resonator (SR) of frequency $\omega_{a}$~\cite{Rocheleau2010,Teufel2011}. The system is placed in a dilution refrigerator, and SR is pumped by carefully spectrally filtered hot and cooled leads. The spectral filtering can be realized by coupling the SR with two microwave resonators, designed at $\omega_{a}$, and $\omega_{-}$ frequencies, respectively. Each resonator is coupled to a copper thin-film resistor, which provides the cold and hot thermal baths in the setup~\cite{Senior2020}. This arrangement can also be extended to an electro-mechanical system in which multiple mechanical resonators are coupled to a microwave circuit~\cite{Massel2012}.

\begin{figure}[t]
\centering
\includegraphics[width=0.50\textwidth]{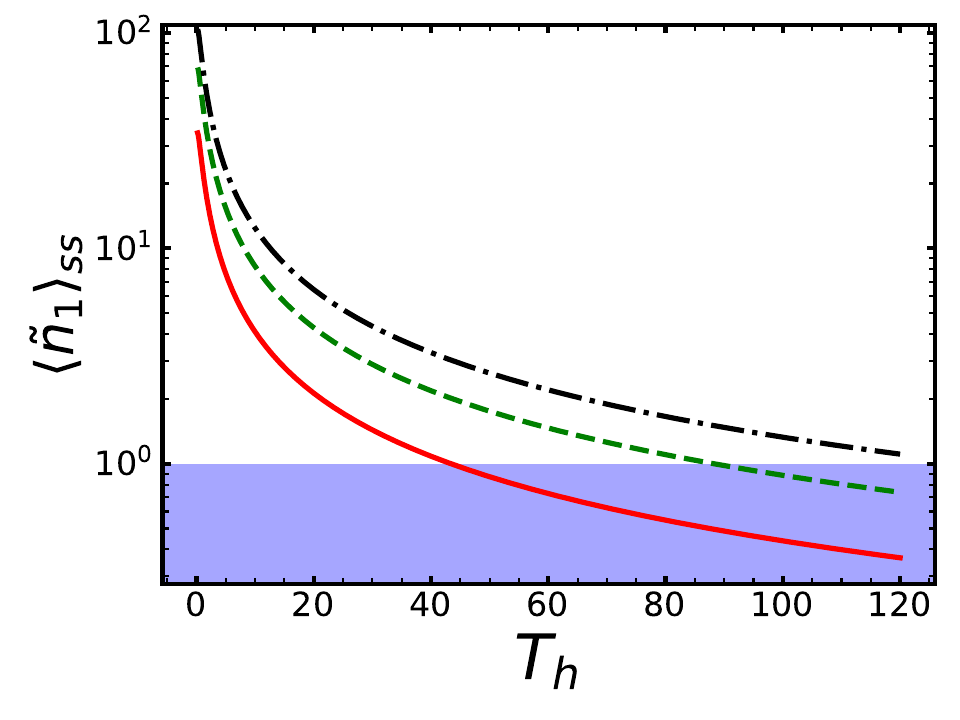}
\caption{\label{fig:8}{\bf Ground-state cooling in case of an electromechanical system.} The steady-state mean phonon number $\langle \tilde{n}_{1}\rangle_{ss}$ as a function of the hot bath temperature $T_{h}$. Solid, dashed, and dot-dashed lines are for $T_{1} \approx 20$, $35$, and $65$ mK, respectively. The other parameters are $\omega_{a} = 2\pi\times 7$ GHz, $\omega_{1} = 2\pi\times 10$ MHz, $g_{1} = 2\pi\times 10$ kHz, $\kappa_{h} = \kappa_{c} = 2\pi\times 1$ MHz, $\kappa_{1} = 2\pi\times 32$ Hz, and $T_{c}\approx 0.34$ K. All the parameters are scaled with the superconducting microwave resonator frequency $\omega_{a} = 2\pi \times 7$ GHz. The hot bath temperature in SI units is $1 \lesssim T_{h}\lesssim 400$ K.}
\end{figure}

For this arrangement, the master equation is given in Eq.~(\ref{eq:filtMEM}) remains valid, and the final mean phonon number is plotted in Fig.~\ref{fig:8}. The results reflect that it is possible to cool a mechanical resonator's motion to its ground-state within the experimentally realizable parameters. We note that the mechanical resonator at a higher temperature (up to a few Kelvins) can be cooled to its ground-state by considering a larger single-photon optomechanical coupling strength. This can be achieved by coupling the microwave circuit to a mechanical resonator via a flux-mediated coupling~\cite{Rodrigues2019}. For the ground-state cooling of a room temperature mechanical resonator, an ultra-strong single-photon coupling $g\sim(\omega_{1},\kappa_{x})$~\cite{PhysRevLett.123.247701} and very high hot bath temperature $T_{h}\sim 10^{9}-10^{10}$ K is required. The ultra-hot temperature of the bath can be challenging to realize experimentally. However, a significant cooling, e.g., $T_\text{eff}\sim 1$ K of a room temperature mechanical resonator, can be achieved within the experimentally available parameters given in Fig.~\ref{fig:8}. Here, $T_\text{eff}$ is the effective temperature of the mechanical resonator. It may be worth emphasizing that the hot bath temperature can be considered higher than the room temperature, as SR is coherently coupled with the cold bath only, which is maintained at a much lower temperature (e.g. $0.1$ K). The SR reaches almost a thermal steady-state at a very low temperature given in Eq.~(\ref{eq:appna}) and verified in Fig.~\ref{fig:4}. We note that, by driving the electrical component of the electro-mechanical system with arbitrary waveform generators that produce noisy electric fields, a hot bath of very high temperature $T_{h}\sim 9500$ K can be engineered~\cite{PhysRevX.7.031044}.
 %%%%%%%%%%%%%%%%%%%%%%%%%%%%%%%%%%%% Figure 8 %%%%%%%%%%%%%%%%%%%%%%%%%%%%%%%%%%%%%%%%%%%%%%%%%%
%%%%%%%%%%%%%%%%%%%%%%%%%%%%%%%%%%%%%%%%%%%%%%%%%%%%%%%%%%%%%%%%%%%%%%%%%%%%%%%%%%%%%%%%%%%%%%%%

(B) A qubit parametrically coupled to a mechanical resonator~\cite{LaHaye2009,Connell2010} can also be considered to implement our scheme (see Fig.~\ref{fig:7}) of cooling. %the mechanical resonators to their quantum ground-state. 
The BSF of cold and hot baths can be realized by coupling two microwave resonators of frequency $\omega_{a}$, and $\omega_{-}$ to the qubit, respectively. The filtered cold and hot bath spectra take Lorentzian form~\cite{PhysRevA.90.032114} centered at frequencies $\omega_{a}$, and $\omega_{-}$, respectively. For this arrangement, the master equation in Eq.~(\ref{eq:filtMEM}) is valid with a difference of replacing the bosonic operators for the cavity mode with the respective Pauli operators~\cite{PhysRevE.90.022102, Naseem_2020}.

(C) In the case of cooling in the optical regime, a suitable candidate for implementing our scheme can be a toroid optical microresonator parametrically coupled to mechanical oscillators~\cite{Schliesser2008}. In this setup, the cold bath can be an electromagnetic vacuum that unavoidably coupled to the optical resonator. The typical spectrum of such a cold bath has a Lorentzian shape~\cite{RevModPhys.86.1391}, and it can be controlled by the linewidth of the optical resonator. An optical fiber taper can be used to couple the hot bath with the optical resonator via a bandpass filter ~\cite{Gelbwaser-Klimovsky2015}. This setup has previously been proposed for the implementation of an autonomous quantum heat engine~\cite{Gelbwaser-Klimovsky2015}. Ground-state cooling of mechanical resonators may not be possible in this setup because of the need for ultra-high temperature thermal bath; however, significant simultaneous cooling of mechanical resonators can be achieved. Alternatively, an ultra-high effective temperature squeezed thermal bath can be employed for the cooling (see Results ``Cooling by heating via a non-thermal bath''). 

(D) Another possible setup in the optical regime can be based on a Fabry--P\'erot cavity as shown in Fig.~\ref{fig:1}. The cold bath can be an electromagnetic vacuum coupled to the cavity mode~\cite{Gelbwaser-Klimovsky2015}, and the hot bath can be realized by a black-body light source successively filtered around frequency $\omega_{-}$.  A bandpass filter is the most straightforward and efficient choice for our purpose to reject the unwanted thermal radiation and select a band centered at a target sideband. Using alternating layers  of thin-film Fabry--P\'erot ``cavities", one can design a filter for a desired wavelength and bandwidth.  The main tuning parameter is the spacing layer's width between reflecting stacks of the Fabry--P\'erot cavities. Besides, the number of the reflecting layers can be used for further tuning, too. We note that the angle of incidence of the input radiation can shift the central wavelength, depending on the blocking layers' refractive index. The additional shift is exploited for fine-tuning in narrowband bandpass filters. Also, at high temperatures, the layers' expansion can cause extra wavelength shifts, which can be significant for our ultra-high-temperature optical bath requirements. We address the question of realization of such a high-temperature bath and its alternatives in (see Results ``Cooling by heating via a non-thermal bath''). Bandwidth is adjusted by an additional dielectric or metal-dielectric blocking layers under a given constraint of the filter's target overall transmission quality. 

Alternatively, it may be possible to simulate the bandpass filtered thermal light using a coherent laser drive at a frequency $\omega_{-}$ with engineered noise corresponding to an
effective high temperature. This practically efficient scheme, however, requires analysis of the master equation and the effective bath correlations functions, which is left open for future studies.

{\bf Dynamics of the system.}
For the optomechanical system shown in Fig.~\ref{fig:1}(a), if we consider a single mechanical resonator of frequency $\omega_{1}$, and bath spectrum filtering of Fig.~\ref{fig:1}(c), the master equation can be written as~\cite{PhysRevE.85.061126}
\begin{eqnarray}
\tilde{ \mathcal{L}}_{C} &=& \gamma_{c}(\tilde{\mathcal{D}}[\tilde{a}]
	+ e^{-\beta_{c}\omega_{a}}\tilde{\mathcal{D}}[\tilde{a}^{\dagger}]) \\ 
\tilde{\mathcal{L}}_{H} &=&\gamma_{h}\zeta_{1}^2\big(\tilde{\mathcal{D}}[\tilde{a}\tilde{b}^{\dagger}_{1}]
	+ e^{-\beta_{h}\omega_{+}}\tilde{\mathcal{D}}[\tilde{a}^{\dagger}\tilde{b}_{1}]\big),
	\\ 
	\tilde{\mathcal{L}}_{1}&=& \gamma_{1}(\tilde{\mathcal{D}}[\tilde{b}_{1}]
	+e^{-\beta_{1}\omega_{1}}\tilde{\mathcal{D}}[\tilde{b}_{1}^{\dagger}]),
\end{eqnarray} 

here $\gamma_{c}, \gamma_{h},$ and $\gamma_{1}$ are relaxation rates which depend on the specific model of the cold, hot and mechanical baths, respectively. In addition, $\beta_{c}, \beta_{h},$ and $\beta_{1}$ are inverse temperatures of the cold, hot and mechanical baths, respectively. In the limit $\beta_{h}\to 0$, the rate equations for the mean number of photons and phonons read
\begin{eqnarray}
\frac{d}{dt}\langle\tilde{n}_{a}\rangle = &-&\gamma_{c}(1-e^{-\beta_{c}\omega_{a}})\langle\tilde{n}_{a}\rangle + \gamma_{c}e^{-\beta_{c}\omega_{a}} \nonumber \\ &+&\gamma_{h}(\langle\tilde{n}_{1}\rangle-\langle\tilde{n}_{a}\rangle),\\
\frac{d}{dt}\langle\tilde{n}_{1}\rangle = &-&\gamma_{1}(1-e^{-\beta_{1}\omega_{1}})\langle\tilde{n}_{1}\rangle + \gamma_{1}e^{-\beta_{1}\omega_{1}} \nonumber \\ &+&\gamma_{h}(\langle\tilde{n}_{a}\rangle-\langle\tilde{n}_{1}\rangle),
\end{eqnarray}
and at steady-state the mean number of quanta in the mechanical resonator take the form
\begin{eqnarray}\label{eq:nan1ss}
\langle\tilde{n}_{1}\rangle_\text{ss} = \frac{\gamma_{c}e^{-\beta_{c}\omega_{a}}+\gamma_{1}e^{-\beta_{1}\omega_{1}}}{-\gamma_{c}e^{-\beta_{c}\omega_{a}}-\gamma_{1}e^{-\beta_{1}\omega_{1}}+\gamma_{c}+\gamma_{1}}.\nonumber \\
\end{eqnarray}
For one-dimensional Ohmic spectral densities of the baths, Eq.~(\ref{eq:nan1ss}) simplifies to
\begin{eqnarray}
\langle\tilde{n}_{1}\rangle_\text{ss} = \frac{\omega_{a}\kappa_{c}\bar{n}_{c}+\omega_{1}\kappa_{1}\bar{n}_{1}}{\omega_{a}\kappa_{c}+\omega_{1}\kappa_{1}},
\end{eqnarray}
which shows that ground-state cooling is possible for the {appropriate system parameters (see Results)}.

\section*{Methods}

{\bf The master equation.}~The free Hamiltonians of the independent thermal baths is given by
\begin{equation}
\hat{H}_{Bx} = \sum_{k,x}\omega_{k,x}\hat{a}^{\dagger}_{k,x}\hat{a}_{k,x},
\end{equation} 
where, $x = H, C, i$ represents the hot, cold, and mechanical baths, respectively. In addition, the infinite number of bath modes are indexed by $k$. We need to attach two baths with the cavity mode, because the hot bath removes energy from the mechanical resonator and dumps it in a bath at lower temperature. In our scheme, cooling with the single optical bath is prohibited by the second law of thermodynamics (cf.~Fig.~\ref{fig:2}).
The interaction of the isolated optomechanical system with the thermal baths is given by the Hamiltonian
\begin{multline}
\hat{H}_{SB}=\sum_k g_{k, \alpha}(\hat{a}+\hat{a}^{\dagger})\otimes(\hat{a}_{k, \alpha}+\hat{a}^{\dagger}_{k, \alpha}) \\ + \sum_{k, i} g_{k,i}(\hat{b}_{i}+\hat{b}_{i}^{\dagger})\otimes(\hat{a}_{k, i}+\hat{a}^{\dagger}_{k,i}), 
\end{multline}
where, $\alpha = H, C$, and the first term describes the interaction of hot and cold baths with the optical mode, and the second term represents the interaction between the mechanical resonator $b_{i}$ and its bath $B_{i}$.

To derive the master equation, we diagonalize the Hamiltonian of the isolated optomechanical system by a unitary transformation
\begin{equation}\label{eq:unitary}
\hat{S} = e^{-\sum_{i}\zeta_{i} \hat{a}^{\dagger}\hat{a}(\hat{b}_{i}^{\dagger}-\hat{b}_{i})},
\end{equation}
where $\zeta_{i} = g_{i}/\omega_{i}$. The diagonalized Hamiltonian reads as
\begin{equation}\label{eq:diagHam}
\tilde{H}_{s} = \omega_{a}\tilde{a}^{\dagger}\tilde{a} + \sum_{i}\big[\omega_{i}\tilde{b}_{i}^{\dagger}\tilde{b}_{i}-\frac{g_{i}^2}{\omega_{i}}(\tilde{a}^{\dagger}\tilde{a})^2\big],
\end{equation}
and the transformed operators are given by
\begin{eqnarray}
\tilde{a} &=& \hat{S}\hat{a}\hat{S}^{\dagger} = \hat{a} e^{-\sum_{i}\zeta_{i}(\hat{b}_{i}^{\dagger}-\hat{b}_{i})},\\ 
\tilde{b}_{i} &=& \hat{S}\hat{b}_{i}\hat{S}^{\dagger} = \hat{b}_{i}-\zeta_{i}\hat{a}^{\dagger}\hat{a}.
\end{eqnarray}
The eigenenergies of the isolated optomechanical system are expressed as
\begin{equation}
E_{n_{a}, m_{i}}= n_{a}\omega_{a}+m_{i}\omega_{{i}}-n^2_{a}\frac{g_{i}^2}{\omega_{i}},
\end{equation} 
where $m_{i}$ is the number of phonons in the mechanical resonator of frequency $\omega_{i}$, and $n_{a}$ is the number of photons in the cavity. The master equation in the interaction picture evaluate to
\begin{eqnarray}
	\label{eq:L_L}
\tilde{ \mathcal{L}}_{\alpha} &=& G_{\alpha}(\omega_{a})\tilde{\mathcal{D}}[\tilde{a}]
	+ G_{\alpha}(-\omega_{a})\tilde{\mathcal{D}}[\tilde{a}^{\dagger}] \\ \nonumber &+& \sum_{i}\zeta_{i}^2\bigg[  G_{\alpha}(\omega_{-})\tilde{\mathcal{D}}[\tilde{a}\tilde{b}^{\dagger}_{i}]
	+ G_{\alpha}(-\omega_{-})\tilde{\mathcal{D}}[\tilde{a}^{\dagger}\tilde{b}_{i}]
	\\ \nonumber &+& G_{\alpha}(\omega_{+})\tilde{\mathcal{D}}[\tilde{a}\tilde{b}_{i}]
	+ G_{\alpha}(-\omega_{+})\tilde{\mathcal{D}}[\tilde{a}^{\dagger}\tilde{b}^{\dagger}_{i}]\bigg],
	\\ 
	\tilde{ \mathcal{L}}_{i}&=& G_{i}(\omega_{i})\tilde{\mathcal{D}}[\tilde{b}_{i}]
	+ G_{i}(-\omega_{i})\tilde{\mathcal{D}}[\tilde{b}_{i}^{\dagger}], \label{eq:L_M}
\end{eqnarray} 
where $\omega_{\pm} = \omega_{a}\pm\omega_{b}$, and the Lindblad dissipators $\tilde{D}[\tilde{o}]$ are defined as
\begin{equation}\label{dissipator}
\tilde{\mathcal{D}}[\tilde{o}] = \frac{1}{2}(2\tilde{o}\tilde{\rho}\tilde{o}^{\dagger}-\tilde{o}^{\dagger}\tilde{o}\tilde{\rho} - \tilde{\rho}\tilde{o}^{\dagger}\tilde{o}),
\end{equation}
and the bath coupling spectrum has the form (we take k$_\text{B}=1$)
\begin{equation}
G_{x}(\omega) = \int^{\infty}_{-\infty}e^{i\omega t}\langle B_{x}(t) B_{x}(0)\rangle dt = e^{\omega/T_{x}}G_{x}(-\omega).
\end{equation}
Note that, in Eq.~(\ref{eq:L_L}), we have considered the weak optomechanical coupling regime~\cite{Gelbwaser-Klimovsky2015}, i.e.,
\begin{equation}
\zeta^2_{i}\langle\tilde{b}_{i}^{\dagger}\tilde{b}_{i}\rangle\ll 1,
\end{equation} 
and ignored all higher order terms $\mathcal{O}(\zeta^{3})$.

\renewcommand\bibsection{\section*{\refname}}
\renewcommand\refname{References}
\bibliographystyle{naturemag}
%\bibliography{main}
\providecommand{\noopsort}[1]{}\providecommand{\singleletter}[1]{#1}%

\end{document}